# The Intellectual Property Protection System of the Foreign Investment Law: Basic Structure, Motivation and Game Logic[*]

Luo Ying[*]

**Abstract**: *The intellectual property protection system constructed by China's Foreign Investment Law has opened a new phase of rule of law protection of intellectual property rights for foreign-invested enterprises, which is an important institutional support indispensable for optimizing the business environment under the rule of law.The development of the regime was influenced by the major concerns of investors' home countries, the "innovation-driven development" strategy, and the trend towards a high level of stringent protection of international intellectual property and investment rules.In addition, there is a latent game of interests between multiple subjects, which can be analyzed by constructing two standard formal game models according to legal game theory.The first game model aims to compare and analyze the gains and losses of China and India's IPR protection system for foreign-invested enterprises to attract foreign investment.The second game model is designed to analyze the benefits of China and foreign investors under their respective possible behaviors before and after the inclusion of IPR protection provisions in the Foreign Investment Law, with the optimal solution being a "moderately cautious" strategy for foreign investors and a "strict enforcement" strategy for China.*

---

[*] Luo Ying (1992-), male, native of Gao'an City, Jiangxi Province, China;A Doctoral degree student at Hunan University Law School , China. (Changsha, Hunan 410082, China; Tel: 18872232314; E-mail:B191900697@hnu.edu.cn, Address: Hunan University School of Law, Changsha, Hunan Province, China).

# Introduction

The issue of intellectual property rights in foreign investment has been a major concern for foreign investors and foreign governments in recent times.Many intellectual property disputes, bilateral investment treaty negotiations, and economic and trade consultations between China and the United States have included intellectual property issues in foreign investment as one of the major negotiation matters.Since 2018, the intensifying "trade war" between China and the United States has also made the protection of intellectual property rights for foreign investment one of the major negotiation matters.How to grasp the scale of intellectual property protection in foreign investment is an important litmus test for the level of foreign investment legislation and the degree of economic openness of the host country.In March 2019, China provided intellectual property protection provisions for the first time in the newly promulgated Foreign Investment Law, thus taking a major step forward in the protection of intellectual property rights of foreign-invested enterprises, which is of milestone significance.This move declares China's firm determination to protect intellectual property rights in foreign investment through explicit legal provisions.However, academic explanations of the reasons and economic logic behind the inclusion of IPR protection provisions in the Foreign Investment Law are somewhat lacking.This paper intends to discuss the significance of the intellectual property protection system of foreign-invested enterprises in building a new system of China's open economy from the perspective of optimizing the business environment.This paper will use normative analysis to explore the domestic and foreign motivations behind the development of the system, and use legal game theory to construct two game models to elucidate the game logic underlying the design of the system.

# I. Optimization of Business Environment and Legal Protection of Intellectual Property Rights of Foreign Invested Enterprises

## (1)The Interaction of Systematic Foreign Opening, Optimization of Business Environment and Intellectual Property Rule Protection

The business environment of a country directly affects the speed and quality of its economic development.Business environment refers to the market economy under the conditions of various types of market business entities to carry out fair competition involving a variety of formal or informal institutional factors formed by the environment.The rule of law is the bottom line and the fundamental guarantee of the business environment. A legalized business environment means a stable, fair, transparent and predictable business environment.Changes in a country's economic development model can profoundly affect the design direction of its legal system.After decades of rapid development, the development mode of China's opening up to the outside world has moved from the policy-oriented opening-up mode of "encouraging exports and introducing foreign investment" to the system-oriented opening-up mode of "all-round, high-level and multi-level" in the new era.The transformation of China's outward opening mode is not only a rational choice in response to the demand for high-level international investment protection competition in the third generation, but also a result of the strong impetus of the urgent need to further enhance the quality of foreign investment utilization.The construction of a new system of institutionalized open economy is inseparable from a high-level, institutionalized, rule-of-law business environment as support.In September 2015, the Central Committee of the Communist Party of China and the State Council issued "Several Opinions on Building a New System of Open Economy," which clearly states, "Building a new system of open economy should keep foreign investment policies stable, transparent and predictable, and create a standardized institutional environment and a stable market environment. Strengthen the rule of law in opening up to the outside world, and focus on building a stable, fair, transparent and predictable business environment."[1]The Fifth Plenary Session of the 19th Central Committee of the Communist Party of China once again stressed the need to "build a new system of a higher level of open economy and raise the level of openness to the outside world".[2]It can be seen that optimizing the business environment is an important element of the system-based external opening development model.

Since 2005, when the World Bank's Doing Business report included the indicator of "intellectual property protection", its impact on the business environment has been widely appreciated by host countries.[3]The latest Doing Business 2020 report continues the tradition of placing emphasis on "intellectual property protection" indicators over the years. It can be said that "intellectual property protection", especially the level of rule of law protection for foreign-invested enterprises, has become an indispensable component for investors or home countries to measure the host country's investment system environment.In this situation, each host country tries to improve its level of IP protection in order to improve its global "Doing Business" ranking and thus gain an advantage in the competition to attract foreign investment.The systemic opening-up model emphasizes adhering to all-round opening-up and dovetailing with advanced international systems and prevailing rules. It means implementing a series of basic legal systems in various fields such as foreign trade, attracting foreign investment, overseas investment and foreign economic and technical cooperation that are connected with the internationally prevailing rules, regulating the practice of opening up to the outside world by means of rules, regulations and laws, and developing institutional competitive advantages.[4]The essence of building a new system of open economy is to promote the formation of an institutional model of open economy development. The institutional open-door development model inherently requires that national economic policies, especially foreign investment policies, should remain stable, continuous, transparent and predictable, and it emphasizes respect for the rule of law and the pursuit of a fair and free market competition environment, which can only be

---

[1] Opinions of the CPC Central Committee and the State Council on Building a New System of Open Economy，http://www.gov.cn/xinwen/2015-09/17/content_2934172.htm[EB/OL].Accessed 1 June 2021.
[2] (Authorized release) Communiqué of the Fifth Plenary Session of the 19th Central Committee of the Communist Party of China，http://www.xinhuanet.com/politics/2020-10/29/c_1126674147.htm[EB/OL].Accessed 1 June 2021.
[3] See Lou Cheng wu，Zhang Guo yong.Research on the Construction of Business Environment Assessment Framework Based on the Subjective Perception of the Market Subject：A Review of Business Environment Assessment Model of the World Bank[J].Contemporary Economic Management,Vol.40, No.6,Jun.2018.
[4] See Xun Kening.Contextual Construction and Path Exploration of China's Institutionalized Foreign Opening[J].Shandong Social Sciences，General Vol.290，No.10,Oct. 2018.

realized under the track of comprehensively promoting the rule of law.[5]

Intellectual property protection system for foreign-invested enterprises is an important part of foreign investment policy.Building an institutional open economy inherently requires that foreign investment policies, including the intellectual property protection system for foreign-invested enterprises, should be modified in line with international institutional standards, while rational choices should be made in the light of actual national conditions.From the current stage, it is the right time for China to strengthen the protection of intellectual property rights for foreign-invested enterprises. It is an important breakthrough in promoting the construction of a new system-based open economy and optimizing the general business environment for competing to attract foreign investment.Therefore, the gradual and orderly docking of international intellectual property protection rules in the field of foreign-invested intellectual property becomes a supporting system to be built for institutional-based opening to the outside world. It can be said that the construction of the intellectual property protection system for foreign-invested enterprises is an indispensable component for optimizing the rule of law business environment. China uses foreign investment law to guarantee the stability, transparency and predictability of foreign investment policies, including intellectual property rights of foreign-invested enterprises, which is conducive to the creation of a general market business environment that abides by the rules, which will further enhance foreign investors' confidence in the Chinese market, thus providing a solid guarantee of the rule of law for the expansion of opening up to the outside world in the new era, upgrading the level of utilization of foreign investment and promoting the construction of a new system of a higher level of open economy.To this end, in March 2019, China provided for the first time for the protection of intellectual property rights in the newly promulgated Foreign Investment Law, thus formally incorporating the intellectual property rights of foreign-invested enterprises into the framework of specialized protection in the form of express legal provisions.

### (2) Intellectual Property Protection Provisions of the Foreign Investment Law: Towards a Basic Framework for    legalization Governance

China's Foreign Investment Law incorporates provisions on intellectual property protection for foreign investment, thus establishing for the first time the basic structure of the intellectual property protection system for foreign-invested enterprises in the form of a basic law on foreign investment.The Foreign Investment Law establishes this institutional structure through Articles 22 and 23.Article 22 of the Foreign Investment Law provides： The State protects the intellectual property rights of foreign investors and foreign-invested enterprises, protects the legitimate rights and interests of intellectual property rights holders and related rights holders, and holds intellectual property rights infringers legally accountable in strict accordance with the law.The State encourages technical cooperation based on the voluntariness principle and commercial rules in the process of foreign investment. The conditions for technical cooperation are determined by equal negotiation between the parties to the investment in accordance with the principle of fairness. Administrative agencies and their staff are prohibited to use administrative means to force any technology transfer.From the interpretive perspective of legal doctrine, the content of intellectual property protection displayed in this article is mainly regulated by law in four aspects: protection of legal rights and interests, strict legal liability for infringement, encouragement of technical cooperation and prohibition of compulsory transfer of technology.This article divides the object of protection of foreign-invested intellectual property into two categories of subjects: one, the protection of intellectual property rights of foreign investors and foreign-invested enterprises. Here the foreign investors also include foreign natural persons, legal persons and unincorporated entities. Second, to protect the legitimate rights and interests of intellectual property rights and related rights holders.It shows that the law protects not only the legitimate rights and interests of the IP rights holders, but also the legitimate rights and interests of the people related to the IP rights. Legitimate rights and interests include not only rights but also economic interests that have not yet formed rights.This article clearly indicates that the legal responsibility of intellectual property infringers should be strictly investigated in accordance with the law, and crack down on intellectual property infringement, reflecting the positive attitude of the state to strictly protect the intellectual property rights of foreign investment.The article also provides that the principle of technical cooperation between China and foreign countries should follow the principle of voluntariness and commercial rules, and encourage the principle of fairness in determining the terms of technical cooperation through equal consultation. This is more consistent with the spirit of the civil law of the civil subjects follow the principle of equality, voluntariness and fairness in carrying out civil activities. This is undoubtedly conducive to the protection of the legitimate rights and interests of both sides of the investment, and is conducive to the Chinese side to carry out technical cooperation through legitimate and legal means to enhance their own technology.In addition, the article also establishes a core principle that prohibits administrative organs and their staff from using administrative means to compel the transfer of technology. This helps to dispel the concerns of foreign investors and foreign governments about the protection of intellectual property rights in China.

---

[5] See Zhong Changbiao.Transformation of China's Opening-up Mode from Chinese Foreign Investment Law[J].Wuhan University Journal(Philosophy & Social Sciences),Vol. 72 ,No. 5, Sept. 2019.

The Foreign Investment Law established the basic principles of the rule of law for the protection of trade secrets of foreign-invested enterprises.Article 23 of the Foreign Investment Law：The administrative organs and their staff shall keep confidential the business secrets known to them, of foreign investors and foreign-invested enterprises during the performance of their duties, and shall not disclose or illegally provide them to others.The article clearly requires the administrative organs and their staff in the process of performing their duties in accordance with the law for the knowledge of foreign investors, foreign-invested enterprises have the obligation of confidentiality. The obligation is not to disclose or illegally provide others with the knowledge of commercial secrets. Administrative organs and their staff in violation of this obligation will be held legally responsible.In summary, the legal framework established by the Foreign Investment Law for the protection of intellectual property rights in foreign investment is as follows: (a) The legal principles for the protection of intellectual property rights of foreign investors and foreign-invested enterprises are established.(b) explicitly prohibit the use of administrative means to compel foreign investors and foreign-invested enterprises to transfer technology; (c) clearly proposed to protect the commercial secrets of foreign investors and foreign-invested enterprises; (d) encourage technical cooperation in the process of foreign investment based on the principle of voluntariness and commercial rules; (e) strictly pursue the legal liability of intellectual property infringers in accordance with the law.The basic legal framework for intellectual property protection built by the Foreign Investment Law is undoubtedly of great institutional support for optimizing the business environment.

## II. The Motivation of the Foreign Investment Law to Construct an Intellectual Property Protection System

### (1) Home countries show significant and long-standing concerns about IPR protection in China

The issue of IPR protection for foreign-invested enterprises in China has long been accompanied by significant concerns of foreign investors and home country governments about meeting the requirements or expectations of capital-exporting countries that trade with China, as well as foreign enterprises seeking investment from China. Most typically, the U.S. government and its investors have shown significant long-term and ongoing concerns about the protection of intellectual property rights in our foreign-invested enterprises. In terms of the early stages of reform and opening up, the momentum of enactment and continuous improvement of most of China's IP laws was largely the result of the influence of U.S. factors.[6] In other words, the process is often accompanied by continued demands under the threat of U.S. trade sanctions.[7] As early as 1979, the U.S.-China Trade Relations Agreement was signed between China and the United States, which obligated the two countries to provide each other with equal protection of intellectual property rights in accordance with international practice. This means that China must begin to learn to study and gradually accept the constraints of a set of rules constructed from the legal systems of major Western industrial powers in areas such as trade and intellectual property. The signing of the U.S.-China Trade Relations Agreement has forced China to be cautious about pressing claims from the U.S. government and foreign investors for greater protection of its intellectual property rights.[8] This was reflected in the subsequent Sino-foreign Joint Venture Law enacted in 1979. In 1988, the United States introduced the Omnibus Trade and Competitiveness Act, which set up the famous "Special 301"[9] Provision to promote adequate and effective protection of U.S. companies' intellectual property rights abroad. Because of China's "deficiencies" in intellectual property protection at the time, the United States placed China on its "priority list" in the first year of "Special 301". In order to improve the status of IPR protection at the time and in response to strong U.S. concerns, China had to work to strengthen IPR protection. As a result of China's efforts, most of the reasons cited by the U.S. in identifying China as a "priority foreign country of concern" were gradually eliminated.[10] In 1989, representatives of the United States and China held the first consultations on the protection of the legitimate rights and interests of domestic investors in intellectual property rights for investment activities in China, and the three U.S.-China negotiations on intellectual property rights in 1994, 1995 and 1996 also revealed the serious concerns of the United States about the protection of intellectual property rights in China.

---

[6] Hu Chonghan.Conflict and Cooperation: The American Factor and China's Modern IPR Legal Process[J].Science Technology and Law Vol.98, No.4, 2012.

[7] Liu G. Patent protection in China is the prerequisite for importation of foreign advanced technology[J]. BMJ,1990,(1).

[8] Hsieh,Pasha L. China-United States Trade Negotiations and Disputes: The WTO and Beyond[J]. Social Science Electronic Publishing,2009,4(2).

[9] Fisher B.UNITED STATES: OMNIBUS TRADE AND COMPETITIVENESS ACT OF 1988 — TITLEI[J]. International Legal Materials,1989,28(1).

[10] Zhou XL.U.S.-China Trade Dispute and China's Intellectual Property RightsProtection.[J].New York University Journal of International Law and Politics,1992,24(3).

Since then, the United States has often initiated WTO dispute settlement procedures on intellectual property issues, particularly the strength and transparency of intellectual property enforcement, even after China's accession to the WTO.For example, in 2010, the U.S. Trade Representative gave the following reason in the Special 301 Report: "China's IPR enforcement system remains largely ineffective and non-deterrent. Widespread IPR violations continue to affect products, brands, and technologies across a wide range of industries, among others."[11]In fact, the standard of IPR protection required by the United States for China at that time was too strict to the extent that it was difficult to enforce. In addition, the U.S. argued that the process of adjudicating cases in China's courts lacked transparency. The reason for this was that fewer cases involving IPR infringement were publicly decided in Chinese courts, and that the judgments in such cases were rarely translated into English, making them difficult to make known to the foreign public.[12]This often makes it difficult for outsiders, especially foreign investors and foreign governments, to get a true picture of the situation, which can subjectively lead to inaccurate or grossly inaccurate news reports.This will inevitably lead to misunderstandings or suspicions about the state of IPR protection in China.[13]At the same time, observers of U.S.-China IPR disputes and some U.S. IP scholars tend to view China's piracy problem in a way that magnifies specific areas of piracy to appeal to public sentiment and thereby lend support to proposed legislative and administrative actions. [14]This often leads to greater misperceptions among the foreign public about the true nature of China's serious piracy problem.

During this period, the United States continued to launch special Section 301 investigations against China on a regular basis. For example, in 2005, the U.S. again placed China on its "priority list of concerns" on the pretext that China had failed to meet its WTO obligations.The U.S. government has asked the Chinese government to curb the frequency of IPR infringement by, but not limited to, increasing the number of criminal prosecutions for IPR infringement, revising and introducing new copyright laws, and adopting the "anti-circumvention" laws proposed in the U.S. Digital Millennium Copyright Act of 1998.[15]Since then, the U.S.-China economic and trade negotiations and the 2012 U.S.-China Bilateral Investment Agreement negotiations have included intellectual property protection issues as the focus of the negotiations.In March 2018, the U.S. filed a dispute settlement proceeding with the WTO against China's alleged forced technology transfer. The U.S. brought China's foreign investment law related to international investment-related intellectual property disputes to the WTO, based on Article 3 national treatment and Article 28 patent rights granted under the TRIPS Agreement.[16]In fact, the escalating U.S.-China trade frictions since 2018 and the EU's continuous suits against China for ineffective IPR protection in international investment arbitration bodies also reflect the great concern shown by foreign investors and foreign countries for IPR and its rights and interests protection.[17] Among them, piracy infringement crackdown, compulsory technology transfer and trade secret protection have been the three key concerns of foreign investors and foreign countries on China's intellectual property protection.[18] China's original "Three Laws on Foreign Investment"[19] have drawn criticism for their long-standing lack of intellectual property protection provisions. It was not until 2019 that China incorporated provisions on the protection of intellectual property rights of foreign investment into the newly introduced Foreign Investment Law, marking a key step towards the rule of law for the protection of intellectual property rights of foreign-invested enterprises. It can be argued that the significant long-term and ongoing concerns of foreign investors and foreign governments are one of the motivations for the move towards rule of law protection of intellectual property rights for foreign-invested enterprises in China.

---

[11] REP OOTU. IIPA's 2010 Special 301Report[EB/OL].2010,IIPA.
https://iipa.org/files/uploads/2018/01/2010_Special_301.pdf. Accessed 1 June 2021.

[12] Volper TE.TRIPS Enforcement in China: A Case for Judicial[J].Brook.j.intl L,2007,(1).

[13] Campbell PE, Pecht M. The Emperor's New Clothes: Intellectual Property Protections in China[J]. Journal of Business & Technology Law,2012,7(1).

[14] See LiHH. Piracy, Prejudice and Profit: A Perspective from US－China Intellectual Property Rights Disputes[J]. Journal of World Intellectual Property,2006,9(6).

[15] McCullagh D. Rampant piracy lands China on 'watch list' _The Bush administration claims China isn't living up to its commitment to a copyright crackdown.[EB/OL].2005,CNET.https://
www.cnet.com/news/rampant-piracy-lands-china-on-watch-list/.Accessed 1 June 2021.

[16] See Zhang Naigen.International Investment Related Intellectual Property and its Dispute Resolution[J].Research on Rule of Law,2020,(1).

**(2) Strongly driven by the trend of high standard and strict protection of international intellectual property**

The rule of law process of IPR protection for foreign-invested enterprises in China has been deeply influenced by the evolution of international IPR protection rules. In 1995, the introduction of the Agreement on Trade-Related Aspects of Intellectual Property Rights (TRIPS) was widely seen as representing a model of legal transplantation aimed at achieving neoliberal globalization. Under this model, the TRIPS agreement does facilitate the process of standardizing IP laws worldwide, regardless of the cultural, economic or political differences between countries.[20] The entry into force of the TRIPS Agreement has also extended intellectual property rights from the domestic private intellectual creation field to the international economic and trade field, and it is the first revolutionary leap in the international intellectual property protection rules, marking a new stage of unified standards in the intellectual property system. The TRIPS Agreement has had a significant and profound impact on the implementation and development of China's intellectual property system today. Since 2001, China's intellectual property legal system has been basically aligned with the TRIPS Agreement after its accession to the WTO. Since 2001, China's IPR legal system has started to be basically consistent with the TRIPS Agreement after its accession to the WTO.

Despite the fact that China's IPR regime generally meets the requirements of almost all key international treaties, the United States and the European Union frequently accuse China of failing to meet its obligations to effectively enforce its IPR laws. At the same time, the U.S. perceives a serious lack of transparency in China's development of IPR rules and access to detailed information on its enforcement in the IPR area. To this end, the United States, Switzerland and Japan have requested detailed information from China regarding its IPR enforcement efforts for the period 2001-2005.[21] They claim that these requests are made under

---

[17] IPR protection inChina rapid,fair,effective[EB/OL].2018,Chinadaily.http://www. chinadaily.com.cn/kindle/2018-12/18/content_37413956.htm.Accessed 2 June 2021.

[18] See SONG Hefa, ZHAO Xing,WU Jingjing.Science & Technology Innovation, Intellectual Property Development Countermeasures under Background of US-China Trade Conflict[J].Bulletin of Chinese Academy of Sciences,Vol. 34. No. 8,2019.

[19] The Law on Chinese-Foreign Cooperative Enterprises, the Law on Business Enterprises and the Law on Foreign Investment Enterprises are together called the "Three Laws on Foreign Investment". At present, the Three Foreign Investment Laws are no longer in legal effect."

[20] See Stoianoff N P.The Influence of the WTO over China's Intellectual Property Regime[J]. The Sydney Law Review,2012,34(1).

[21] See Kanji,O.Paper Dragon: Inadequate Protection of Intellectual Property Rights in China[J]. Michigan Journal

Article 63.3 of the TRIPS Agreement:"Each Member shall be prepared to supply, in response to a written request from another Member, information of the sort referred to in paragraph 1. A Member, having reason to believe that a specific judicial decision or administrative ruling or bilateral agreement in the area of intellectual property rights affects its rights under this Agreement, may also request in writing to be given access to or be informed in sufficient detail of such specific judicial decisions or administrative rulings or bilateral agreements."[22]In the Special 301 Report issued in 2015, the United States concluded that China still has significant deficiencies in protecting trade secrets, and that trade secret misappropriation and trademark counterfeiting remain quite serious.[23]For more than a decade after the TRIPS Agreement came into force, the United States and other developed Western countries have been dissatisfied with the level of IPR protection established by the Agreement, and therefore conceived of an IPR enforcement agreement that exceeds the TRIPS Agreement standards.This is because the IPR protection standards in the existing TRIPS Agreement can hardly meet the needs of developed countries to compete for a new round of IPR supremacy, so developed countries are trying their best to promote a high level of IPR protection standards, resulting in the international IPR protection in the multilateral framework towards the "super TRIPS" IPR protection rules.[24]This is reflected in the increasingly broad and specific content of IPR provisions in plurilateral, regional and bilateral agreements, and in the significant increase in protection standards and enforcement.[25]It is against this background that the intellectual property rights of foreign-invested enterprises in China have accelerated into the path of rule of law protection.

**(3) A rational choice to comply with the high level of protection required by international investment rules**

The rule of law in the protection of intellectual property rights of foreign-invested enterprises in China is heavily influenced not only by international intellectual property rules, but also by the development of international investment rules.International investment treaties themselves do not address the rules of IPR protection, and their protection of IPR is mainly achieved by invoking the relevant IPR treaties.[26]International investment treaties usually include IPRs in the scope of "investment" in their investment treatment provisions, thus indirectly introducing and strengthening IPR rules and empowering IPR owners to use the investment dispute settlement mechanism to initiate investment arbitration proceedings against host countries.The Morris case and the Eli Lilly v. Canada case show that IPR policies of host countries are being profoundly affected by the provisions of international investment treaties that give investors a mechanism to arbitrate IPR issues against the host country through the ISDS.[27]In recent times, international investment rules have developed towards a high level of investment

---

of International Law,2006,27(4).

[22] Agreementon Trade-Related Aspects of Intellectual Property Rights[EB/OL]. https://www.wto.org/english/docs_e/legal_e/27-trips_01_e.htm.Accessed 2 June 2021.

[23] 2015 SPECIAL 301 REPORTONCOPYRIGHT PROTECTIONANDENFORCEMENT[EB/OL].2015,IIPA. https://www.iipa.org/files/uploads/2018/01/2015_Special_301.pdf.Accessed 2 June 2021.

[24] Sell SK.TRIPS Was Never Enough: VerticalForum Shifting,FTAs,ACTA,and TPP[J].Journal of Intellectual Property Law,2011,18(2).

[25] See James Boyle, The Second Enclosure Movement and the Construction of the Public Domain[J].66 Law & Contemp. Probs, No.37, 2003.

[26] See Zhang Jianbang .Modern Transformation of the IPR Protection Regime in International Investment Treaties[J].China Legal Science, No.4,2013.

[27] See Xu Shu.The dilemma of intellectual property protection under international investment treaties and its response[J].Law Science,Issue 5, 2019.

liberalization and facilitation, and the system of granting pre-entry national treatment to foreign-invested enterprises has become a powerful trend, leading to a great change in the tendency of international investment legislation.

At the same time, with the advent of the fourth "industrial revolution" represented by 5G technology, quantum communication and artificial intelligence, the rules of IP protection in the international investment field are also undergoing profound changes.For example, the U.S. Model Bilateral Investment Agreement, the TPP Agreement, and the Anti-Counterfeiting Trade Agreement (ACTA) all impose obligations on members that go beyond the IP protection and enforcement obligations under the TRIPS Agreement.[28] For example, the United States has proposed high standards of IPR protection for China in the U.S.-China IPR negotiations, the U.S.-China trade negotiations, and the U.S.-China Model Bilateral Investment Agreement (2012).[29] Developing countries have been forced by this trend to respond to this new high standard and increasingly stringent IPR protection rules.It can be seen that the high standard and the tendency towards stringent international investment and IPR protection rules are another important factor that pushed China to incorporate IPR protection provisions in foreign investment into the Foreign Investment Law.

**(4) The inevitable result of deepening the "innovation-driven development" strategy**

Behind the transformation of the national economic development strategy lies the real need for economic and social development changes, and this need will determine the direction of the relevant legal system design.In 2002, China ranked 43rd in the global innovation capacity and 31st in overall international competitiveness, which is in the lower middle position.[30] After years of development, China has made great progress in building its innovation capacity. As of 2015, China has ranked 19th in the global innovation index and 25th in the index of intellectual property protection strength, indicating that the gap between China's innovation capacity and innovative countries has been narrowing.[31] Despite this, in fact, most Chinese industries are still in the middle and low end of the global value chain, some key technologies are still not self-sufficient, piracy and infringement are still quite serious, and the quality and conversion rate of patents are still low. There is an urgent need to improve the system of incentive innovation from the legal system level, so as to better stimulate the vitality of innovation and creativity of the whole society.Therefore, in order to improve the above situation and effectively respond to the complex domestic and international development situation brought about by the evolution of the new generation of international investment rules, the communiqué of the 18th National Congress of the Communist Party of China stated, "Adhere to the road of independent innovation with Chinese characteristics and implement the innovation-driven development strategy." Thus, innovation-driven development has been elevated to a national strategy.

In contemporary society, the intellectual property system has become the basic guarantee system to stimulate innovation. IPR strategy is in an important position in the innovation-driven development strategy. The degree of intellectual property protection of a country depends on its own economic development level.As China's economic development moves into a new stage of high-quality development, its level of intellectual property protection has also been improved.In 2016, China released the Outline of the National Innovation-driven Development Strategy, which states, "We should build a national innovation system, improve the policy system to stimulate innovation and the legal system to protect innovation. Deepen the implementation of the action plan of intellectual property strategy, and improve the ability to create, use, protect and manage intellectual property rights. Enhance the awareness of intellectual property protection for all people and strengthen the role of the intellectual property system as a basic safeguard for innovation."[32]The basic idea is to rely on the "two wheels" of scientific and technological innovation and institutional innovation to coordinate and sustain each other, to form a systematic capacity for continuous innovation, to adjust all institutional mechanisms that do not adapt to innovation-driven development, and to build a national innovation system to maximize the release of innovative vitality.The 19th National Congress of the Communist Party of China also clearly proposed that "we should firmly implement the innovation-driven development strategy, accelerate the construction of an innovative country, and strengthen the creation, protection and application of intellectual property rights."[33]In fact, the "innovation-driven development" strategy can be regarded as an "upgraded" version of the "strong innovation country" and "strong intellectual property rights" strategies, the basic requirement of which is to make the legal policies on stimulating innovation, especially innovation in the field of intellectual property rights, present a long-term mechanism of rule of law.

---

[28] Ma Zhong-fa.Status, Evolution and Characteristics of the International Intellectual Property Legal System[J].Journal of Anhui Normal University(Hum.& Soc.Sci.),Vol.46,No.3,2018.
[29] See Chen Rywt. China's International Investment and the United States[J]. The Journal of Corporate Accounting & Finance,2015,26(6).
[30] Wang Yan.Authoritative report: China's overall competitiveness ranking rises to 31st in the world[EB/OL].https://www.chinanews.com/2002-12-05/26/250581.html.Accessed 2 June 2021.
[31] National Innovation Index Shows Steady Improvement in China's Innovation Capability[EB/OL].http://big5.xinhuanet.com/gate/big5/www.xinhuanet.com/politics/2015-07/09/c_128000197.htm.Accessed 2 June 2021.

  China's implementation of the "innovation-driven development" strategy is not only a practical need to improve the quality and level of domestic economic development, but also an important measure to adapt to the mainstream international intellectual property protection standards in order to "force" the domestic accelerated pace of independent innovation. In 2018, China's new industries, new industries, new business models and other "three new" economic value added was 14,536.9 billion yuan, equivalent to 16.1% of GDP.[34]This is actually inseparable from the in-depth implementation of the "innovation-driven development" strategy. The implementation of the "innovation-driven development" strategy implies the need to abandon both the previous high-resource-consuming and sloppy economic development model and the traditional model of economic development that relies entirely on investment, exports and consumption.It focuses on knowledge innovation, technological innovation and institutional innovation to accelerate the formation of a positive environment to stimulate innovation development, so as to achieve a major transformation from "Made in China" to "Created in China".This requires the design or amendment of the domestic IPR legal system to be oriented to stimulate independent innovation, while balancing the relationship between knowledge exclusivity and knowledge sharing, in order to achieve stimulating innovation

---

[32]  The Central Committee of the Communist Party of China State Council issued the "National Innovation-driven Development Strategy Outline[EB/OL].http://www.gov.cn/xinwen/2016-05/19/content_5074812.htm.Accessed 3 June 2021.

[33]  Xi Jinping: Deciding to build a well-off society across the board and seizing the great victory of socialism with Chinese characteristics in the new era - Report at the 19th National Congress of the Communist Party of China[EB/OL].http://www.gov.cn/zhuanti/2017-10/27/content_5234876.htm.Accessed 3 June 2021.

[34]  Innovation-driven development strategy to help economic growth[EB/OL].http://news.china.com.cn/live/2019-10/23/content_580839.htm.Accessed 3 June 2021.

development.Under the strategy of "innovation-driven development", the rule of law of intellectual property protection in the field of foreign investment should also be oriented to stimulate independent innovation.This means that the intellectual property rights of foreign-invested enterprises must be strictly protected, a regular institutional protection mechanism must be formed, and the predictability of intellectual property rights enforcement and justice must be strengthened, so as to create an "upgraded" version of intellectual property rights protection in the field of foreign investment.[35]Therefore, the in-depth implementation of the "innovation-driven development" strategy is the internal reason for the Foreign Investment Law to incorporate the intellectual property rights of foreign-invested enterprises into the protection of the rule of law.

## III. Business Game Analysis:The Game Logic of the Foreign Investment Law in Constructing the Intellectual Property Protection System for Foreign Invested Enterprises

The design of the legal system is often based on the logic of rational economic games. This is also the case with the design of China's intellectual property protection system for foreign-invested enterprises. From the perspective of legal game theory, China has constructed a basic legal system structure for the protection of intellectual property rights of foreign-invested enterprises in the newly introduced Foreign Investment Law, behind which lies a complex game process of interests.In this regard, we try to construct two game models to analyze them.The first model is mainly constructed from the perspective of providing "legislative protection" for IPRs of foreign-invested enterprises, which is a form of legislative protection;The second model is mainly constructed from the perspective of providing "enforcement protection" for IPRs of foreign-invested enterprises, which is a substantive protection in terms of enforcement.The author argues that the combination of the two models helps to control variables to simplify the difficulty of understanding and more comprehensively elaborate the game logic of the rule of law protection of intellectual property rights for foreign-invested enterprises in China. After all, the rule of law of IPR protection for foreign-invested enterprises in China cannot be separated from both formal protection at the legislative level and substantive protection at the enforcement level.Only in this way can the intellectual property provisions incorporated in the Foreign Investment Law truly play an important role in optimizing the business environment and enhancing the use of foreign investment.The premise behind the application of game theory to the analysis of such problems is the agreement that game theory can provide insight to those who wish to understand how law affects the behavior of the subjects involved.

### (1) The first standard form game: the game of attracting foreign investment between China and foreign countries

The first game model that needs to be constructed is the standard form game, which is sometimes referred to as the strategic form game.The standard form game consists of three elements: 1. the participants in the game; 2. the strategic behaviors that the participants may implement; and 3. the benefits for the participants under each possible combination of

---

[35] See Kong Xiangjun.Globalization, Innovation-Driven Development and the Upgrade of IP Rule of Law[J].Journal of Law Application,Issue 1, 2014.

behaviors.The first standard form game model is used to model the game process in which China and foreign countries compete to attract foreign investment using different IPR protection regimes for foreign-invested enterprises.

In order to simplify the model to highlight the main points, India can be tried to be used as a typical representative of other countries. The author chooses India as a substitute for other countries to analyze the game of attracting foreign investment between China and foreign countries mainly based on the following reasons: First, in general, the investment environment of major developed countries in Europe and America has not changed much in the past decade or so, and their foreign-related legal environment has been quite mature and perfect.In particular, they have more limited room for improvement regarding the standards of IPR protection in foreign investment, thus leading to little fluctuation in the attractiveness of the regime for foreign investment.This model is intended to illustrate the impact of the foreign investment IPR protection system implemented by the host country on the competition to attract foreign investment, and therefore it is appropriate to choose the host country with some room for improvement in this system as the game participant. India is suitable because there is still some room for improvement in the IPR protection system for foreign investment compared to the major developed countries in Europe and the US.Second, as the most promising developing countries, the BRICS countries represent the development level of the world's emerging markets and are in a more advantageous development position among all developing countries.Except for the BRICS countries, other developing countries are still more limited in terms of market size, economic development potential, and investment system environment in terms of attractiveness to foreign capital.In particular, there is too much room for improvement in their IPR protection system for foreign investment to attract sufficient foreign investment.This has resulted in these developing countries not competing with China enough to attract foreign investment and in total, unlike India as one of the BRICS countries.Third, among the BRICS countries, except for China, India has certain advantages in terms of population market size, total economic development, level of foreign-related intellectual property protection, and attractiveness to foreign investment compared to other BRICS countries (such as Brazil, Russia, and South Africa) in general.It means that India's IPR regime for foreign investment has some conditions to compete with China in attracting foreign investment.According to the official website of BRICS 2019, the United Nations Conference on Trade and Development report and the World Investment Report data： India's total GDP reached $2.94 trillion, attracting total foreign investment of about $49.98 billion; Brazil's total GDP reached $1.62 trillion, attracting total foreign investment of about $75 billion; Russia's total GDP reached $1.69 trillion, attracting total foreign investment of about $21 billion; South Africa's total GDP reached $351.36 billion, attracting total foreign investment The total GDP of South Africa reached USD 351.36 billion, attracting a total foreign investment of about USD 4.6 billion.In summary, from the perspective of economic volume, room for improvement of foreign-related IPR protection system and attractiveness to foreign investors, India is not the only choice, but it is generally more suitable.

It is easy to determine that the first element of this standard form game model, i.e., the game participants, is China and India. The possible strategic behavior of the second element, i.e., the participants, is less easy to determine than that of the first element.Determining the spatial scope of strategic behavior is the most important element of this standard form game. There are various strategies used by investment host countries to attract FDI, but the scope of the strategic behavior chosen by China and India as investment host countries can be appropriately limited in this model.This is because the range of possible strategic behaviors of the game participants depends on the purpose for which we construct this model. This game model aims to analyze the impact of the IPR protection regime on attracting foreign investment in the host country's foreign-invested enterprises.At present, since the concept of intellectual property protection has been recognized in the host countries of investment, there are almost no countries that do not protect the intellectual property rights of foreign-invested enterprises at all.Therefore, we can judge that there are two kinds of IPR strategic behaviors of foreign-invested enterprises in China and India: one is "moderate protection", which refers to IPR protection in the form of foreign investment protection commitments or policies; the other is "strict protection", which refers to IPR protection in the form of a clear foreign investment legal system.Specifically, it can be divided into the following cases: first, both China and India choose "moderate protection"; second, China chooses "strict protection" and India chooses "moderate protection"; third, China chooses "moderate protection" and India chooses "strict protection"; fourth, both China and India choose "strict protection". "Third, China chooses "moderate protection" and India chooses "strict protection"; fourth, both China and India choose "strict protection".

The third element is more critical and requires exploring the structure of possible gains for China and India under each possible combination of strategic behavior. In this model, we consider the IPR protection regime for foreign-invested firms as a means of attracting foreign investment in the host country.Assuming that China and India need to compete to attract a total of $300 billion in foreign investment, the degree of improvement of foreign intellectual property protection system is one of the factors affecting the entry of foreign investment.If all other conditions are equal between China and India, it is assumed that a strategy of "moderate protection" of foreign intellectual property rights will attract 40% of foreign investment and lead to 10% of profitable investment.According to economic principles, every 1% increase in investment can contribute 0.2% to economic growth. Therefore, the proportion of foreign investment that can contribute to the economic growth of the host country is 20% of the total foreign investment attracted.The tax benefit is calculated based on the host country corporate income tax rate, which is known to be generally 20% in India and 25% in China in 2019, with legislative and enforcement costs of $1 billion.Similarly, assuming that the adoption of the strategy of "strict protection" of intellectual property rights of foreign-invested enterprises can attract 60% of foreign investment, the rate of profitability of the investment is also 10%, the proportion of foreign investment that can contribute to the economic growth of the host country is also 20% of the total foreign investment attracted, and the tax revenue is also calculated at the corporate income tax rate of the host country, the legislative and enforcement costs are $2 billion.

In 2019, for example, if both India and China adopt the "moderate protection" strategy, the cost to India is assumed to be $1 billion (legislative and enforcement costs), and the benefit consists of two parts: one part of the benefit is the tax on the profitability of the foreign investment in India, which is calculated as 3,00 × 40% × 10% × 20% = $2.4 billion; the other part of the benefit is the total GDP growth brought to India by the foreign investment in the year, which is 300 x 40% x 20% = $24 billion.Therefore, the final total benefit to India is revenue minus costs can be obtained as 24 + 2.4 - 1 = $25.4 billion.In the same way as this calculation, the cost to China is also $1 billion (legislative and enforcement costs), and the benefit also includes two

parts: one part of the benefit is the tax on the investment profit brought to China by foreign capital, which is the same as India, calculated as 300 x 40% x 10% x 25% = $3 billion; the other part of the benefit is the total GDP growth brought to China by foreign capital in that year The other part of the gain is the total GDP growth brought by foreign capital to China in that year, which can be calculated as 300×40%×20%=24 billion USD.Thus, the final total gain for China is 24 + 3 - 1 = $26 billion.Similarly, if both India and China adopt a "strict protection" strategy, the cost to India is $2 billion (legislative and enforcement costs), and the benefit calculated according to the above formula is 300 x 60% x 10% x 20% + 300 x 60% x 20% = $39.6 billion.Thus, India's final total gain is 39.6 - 2 = $37.6 billion; China's cost is also $2 billion (legislative and enforcement costs), and its gain yields 300 x 60% x 10% x 25% + 300 x 60% x 20% = $40.5 billion. Thus, China's final total gain is $40.5 - 2 = $38.5 billion.Using the same calculation method as above, if India uses "moderate protection" and China uses "strict protection," the cost to India is $1 billion and the final total benefit is $25.4 billion; the cost to China is $2 billion and the final total benefit is $38.5 billion. China's cost is $2 billion, and the total gain is $38.5 billion. If India adopts "strict protection" and China adopts "moderate protection". At this point, India's cost is $2 billion, and the final total benefit is $37.6 billion; China's cost is $1 billion, and the final total benefit is $38.5 billion.

Its benefits and costs can be represented using a binary matrix (Figure 1). By convention, the first gain value in each cell is the gain of the participant located in the row position, and the second gain is the gain of the participant located in the column position.In the figure we can arbitrarily assume that India is the participant in the row position and China is the participant in the column position, so that India is the first gain in each cell and China is the second gain in each cell.

|  |  | China | |
|---|---|---|---|
|  |  | Moderate protection | strict protection |
| India | Moderate protection | $25.4 billion, $26 billion | $25.4 billion, $38.5 billion |
|  | strict protection | $37.6 billion, $26 billion | $37.6 billion, $38.5 billion |

(Figure 1: Binary matrix model of the game of attracting foreign investment between China and India)

From Figure 1, we can see that the optimal solution of this game model for China is for India to adopt a "moderate protection" strategy and for China to adopt a "strict protection" strategy, which gives China the greatest benefit.As of the end of 2019, the reality is that India adopts a "moderate protection" strategy for the protection of intellectual property rights of foreign-invested enterprises.Therefore, for China, the best strategic choice should be "strict protection". It is not difficult to understand, then, that in 2019, China will include provisions in the Foreign Investment Law for "strict protection" of intellectual property rights.

(2) **The second standard form game: the game between China and foreign-invested enterprises to obtain IPR benefits**

The second game model that needs to be constructed arises between foreign investors and

China. This game model still consists of three basic elements: the game participants, the possible strategic behavior, and the payoffs under each possible strategy.This model is intended to illustrate the benefits to China and foreign investors as participants in the game under their respective possible behaviors before and after the inclusion of IPR protection provisions in the Foreign Investment Law, so as to arrive at the optimal choice for China.In this model, the first element of the game is the Chinese and foreign investors as participants in the game. The second element of the game is still the possible strategic behavior of the participants. China as a participant in the game may have two strategies: "strict enforcement" and "moderate enforcement".The term "strict enforcement" refers specifically to China's strict enforcement after the inclusion of IPR protection provisions in the Foreign Investment Law, while the term "moderate enforcement" refers specifically to China's moderate enforcement before the inclusion of IPR protection provisions in the Foreign Investment Law.Foreign investors, as the other side of the game, also have two strategies, one is "very careful" to protect their intellectual property rights, and the other is "moderately careful" to protect their intellectual property rights.The term "very cautious" refers to the fact that prior to the inclusion of IPR provisions in China's Foreign Investment Law, foreign investors were required to spend more specifically to protect their IPR from infringement and piracy. "Moderately cautious" refers to the fact that after the inclusion of IPR provisions in China's Foreign Investment Law, foreign investors can reduce costs to protect their IPR from infringement and piracy.The third element of the game model that needs to be determined now is the respective benefit and cost profiles of the game participants under the possible implementation of the strategies.To determine the benefits and costs of the game participants, several reference factors must first be set and the corresponding variables need to be controlled.It is known that China attracted a total of $134.97 billion in foreign investment in 2018.[36]In 2019, China will attract a total of about $141.23 billion in foreign investment.[37]As a result, China attracted a total of about $6.26 billion more foreign investment in 2019 than in 2018.In addition, it should be noted that the following hypothetical data in this game model are mainly based on the data from the 2019 State of Intellectual Property Protection in China released by the State Intellectual Property Office of China, but the data from the report are not directly quoted in the model. The author has deliberately made simplifications to facilitate people's understanding of the key points in the model and the basic logic of the game.

If China adopts a "strict enforcement" strategy, its administrative enforcement costs are assumed to be $2 billion.If the total size of the gains from the "copycat economy" before the adoption of "strict enforcement" was $20 billion, the loss of gains from the reduction of "copycat (piracy)" behavior and the risk of penalties for some Chinese investors under "strict enforcement" would be $4 billion This part of the loss of revenue needs to be included in the cost of China, because although "copycat" behavior is not legal, but objectively promote the development of the "copycat economy", crackdown on "copycat" behavior will bring some pressure on the domestic part of the economic growth.In addition, China's crackdown on piracy will also detract from tax revenues.If the corporate income tax is calculated at 25%, the tax loss would be 40 x

---

[36] Record high China attracts foreign investment of 885.61 billion yuan in 2018[EB/OL].http://www.gov.cn/xinwen/2019-01/14/content_5357799.htm.Accessed 4 June 2021.

[37] China's actual use of foreign investment in 2019 is $141.23 billion[EB/OL].http://news.cctv.com/2020/11/06/ARTIMEw884VJ5V6rqZehw4DN201106.shtml.Accessed 4 June 2021.

25% = $1 billion. Therefore, the total cost under China's "strict enforcement" would be $2+4+1=$7 billion.So what are the gains from China's "strict enforcement" strategy? First, part of China's gain is that the total amount of foreign investment attracted increased by about $141.23 - 134.97 = $6.26 billion. This is an increase of 62.6 ÷ 1349.7 = 4.64% compared to 2018.According to the principle of economics, every 1% increase in investment can boost the economy by 0.2%. Therefore, the total amount of foreign investment attracted by this part can also drive economic growth is 62.6 × 0.2=1.252 billion dollars.Second, based on the average profitability of Chinese enterprises, the profitability of investment in real enterprises is generally between 10% and 12%. Assuming a 10% profit margin for foreign investment, based on a 25% corporate income tax, Chinese tax revenue could be an additional 62.6 x 10% x 25% = $0.1565 billion.Again, China's crackdown on "copycat" behavior will "force" domestic investors to increase investment in research and development to strengthen independent innovation in scientific research. Assuming that the total amount of money invested in R&D is $20 billion, the investment profit rate is also 10%, and the proportion of economic growth is still calculated in accordance with the above-mentioned economic principles, the investment profit is $20 × 10%=$2 billion, and the economic growth is $20 × 0.2=$4 billion.In addition, according to China's tax regulations, profits from investment in scientific and technological research and development are entitled to a preferential tax rate of 15%. Therefore, the tax income from the investment profit can be calculated at 15% of the corporate income tax. This results in a tax revenue of 20 x 10% x 15% = USD 0.3billion.Finally, one can assume a benefit of $5 billion in patent licensing fees saved from scientific research and development. Overall, then, the total benefit to China is 1.252+0.1565+20+2+4+0.3+5=$32.7085 billion.Excluding the total cost of the preceding, China's final total benefit is $32.7085 - 7 = $25.708 billion.

If China adopts a "moderate enforcement" strategy, its legislative and enforcement costs are assumed to be $1 billion, the total derogation from attracting foreign investment is $2 billion, and Chinese domestic investors pay $10 billion in patent royalties to foreign investors.Therefore, the total cost for China under "moderate enforcement" would be $1+2+10=$13 billion.So, what are the benefits to China under this strategy? First, China can save $1 billion in enforcement costs; second, China can capture the benefits of the "copycat economy" for domestic investors, still assuming that the total benefits are $20 billion; third, the tax benefits of the "copycat economy" are Third, the tax benefit from the "cottage industry" is $20×25% = $5 billion; and fourth, domestic investors pay $5 billion less in royalties due to the "copycat economy".At this point, China's total gain is 1 + 20+ 5 + 5 = $31 billion, and excluding costs, the final gain can be $31 - 13 = $18 billion.The scenarios of benefits and costs under the "very cautious" and "moderately cautious" strategies of foreign investors are relatively simple compared to China.If a foreign investor adopts a "very cautious" strategy to protect its IP, it is assumed to spend $10 billion in litigation costs and IPR maintenance costs and $45 billion in IPR revenues and litigation proceeds.[38]Its final total revenue is $45 - 10 = $35 billion.If a foreign investor adopts a "moderately cautious" strategy to protect its IP, it is assumed to spend $5 billion in litigation costs and IPR maintenance costs, and $30 billion in IPR revenues and litigation gains.Its final total revenue is $35-5= $30 billion.The entire gain/loss scenario can be represented using a binary matrix (Figure 2). Still by convention, the first gain value in each cell is the gain of the participant located in the row position, and the second gain is the gain of the participant located in the column position.In the figure, we can arbitrarily assume that the foreign investor is the participant in the row position and China is the participant in the column position. Thus, the foreign investor is the first gain in each cell, and China is the second gain in each cell.

---

[38] Generally speaking, the amount of damages for IPR infringement is determined by applying, in order, the actual loss suffered by the IPR owner due to infringement and the benefit gained by the infringer due to infringement. If punitive damages for IPR infringement are applied, the punitive damages for IPR infringement are about 3 times of the loss suffered due to infringement, while the proceeds developed from IPR exceed 5 times more than 5 times, if punitive damages are applied, the lawsuit will also generate part of the proceeds. Such a ratio of 1:4.5 is designed here to make a conservative estimate, mainly for the convenience of calculation.

|  |  | China | |
|---|---|---|---|
|  |  | strict enforcement | moderate enforcement |
| Foreign Investors | very careful | 35，25.7085 | 35, 18 |
|  | moderately careful | 30, 18 | 30, 25.7085 |

( Monetary unit of measure: US$ billion)

(Figure 2: A binary matrix model of the game between China and foreign-invested firms acquiring IPR gains)

From Figure 2, we can see that for China, the optimal solution of this game model is for foreign investors to adopt a "moderately cautious" strategy and for China to adopt a "strict enforcement" strategy. At this point, China can obtain the maximum benefit.Therefore, the best strategic choice for China should be "strict enforcement" to protect the intellectual property rights of foreign investors.In 2019, China included provisions on intellectual property protection in the Foreign Investment Law, thus demonstrating its strong political and legal determination to strictly protect the intellectual property rights of foreign investors. This strategic choice will do more good than harm to China in terms of its economic benefits.

# Conclusion

The inclusion of intellectual property provisions in the Foreign Investment Law is an important beginning of the move towards multi-level and all-round protection of intellectual property rights of foreign-invested enterprises in China under the rule of law.There are complex reasons behind China's inclusion of intellectual property protection provisions in the new basic law on foreign investment, which is not only a rational choice to cope with external international competitive pressure, but also the result of the endogenous impetus under the domestic strategy of promoting the implementation of "innovation-driven development".The relevant model constructed in this paper based on legal game theory provides a visual and intuitive representation for a deep understanding of the economic logic behind the rule of law of intellectual property protection for foreign-invested enterprises.